\RequirePackage{fix-cm}
\RequirePackage{graphicx}
\RequirePackage{amsmath,amssymb}
\RequirePackage{mathptmx}
\RequirePackage{isotope}
\RequirePackage{cite}
\RequirePackage{hyperref}
\RequirePackage{booktabs}
\RequirePackage{siunitx}

\documentclass[10pt]{iopart}

\usepackage{upgreek}
\usepackage{color}
\usepackage{xcolor}
\usepackage{doi}
\usepackage[symbol*]{footmisc}
\usepackage{lineno}
\usepackage{booktabs}

\begin{document}
\title[]{A novel ppm-precise absolute calibration method for precision high-voltage dividers}

\author{O. Rest$^1$, D. Winzen$^1$, S. Bauer$^2$, R. Berendes$^1$, J. Meisner$^2$, T. Th\"ummler$^3$, S. W\"ustling$^4$, C. Weinheimer$^1$}

\address{$^1$ Institut f\"ur Kernphysik, Westf\"alische Wilhelms-Universität M\"unster, Wilhelm-Klemm-Str. 9, 48149 M\"unster, Germany}

\address{$^2$ Physikalisch-Technische Bundesanstalt, Bundesallee 100 , 38116 Braunschweig, Germany}

\address{$^3$ Karlsruhe Institute of Technology~(KIT), Institute for Nuclear Physics~(IKP), Hermann-von-Helmholtz-Platz 1, 76344 Eggenstein-Leopoldshafen, Germany}

\address{$^4$ Karlsruhe Institute of Technology~(KIT), Institute for Data Processing and Electronics~(IPE), Hermann-von-Helmholtz-Platz 1, 76344 Eggenstein-Leopoldshafen, Germany}

\ead{o.rest@uni-muenster.de}
\vspace{10pt}
\begin{indented}
\item[]February 2019
\end{indented}

\begin{abstract}
The most common method to measure direct current high voltage (HV) down to the ppm-level is to use resistive high-voltage dividers. Such devices scale the HV into a range where it can be compared with precision digital voltmeters to reference voltages sources, which can be traced back to Josephson voltage standards. So far the calibration of the scale factors of HV dividers for voltages above 1~kV could only be done at metrology institutes and sometimes involves round-robin tests among several institutions to get reliable results. Here we present a novel absolute calibration method based on the measurement of a differential scale factor, which can be performed with commercial equipment and outside metrology institutes. We demonstrate that reproducible measurements up to 35~kV can be performed with relative uncertainties below $1\cdot10^{-6}$. This method is not restricted to metrology institutes and offers the possibility to determine the linearity of high-voltage dividers for a wide range of applications.
\end{abstract}

%
%
%
%
\ioptwocol

\section{Introduction}
\label{section_intro}
Precision measurements of direct current (DC) high voltage (HV) are important for many applications in physics, e.g. to record an integral spectrum of tritium-$\upbeta$-electrons with the KATRIN neutrino mass experiment \cite{Drexlin:2013lha} or for determining kinetic energies of electrons with electron coolers at ion storage rings \cite{ullmann}. The scope of applications is not limited to fundamental research, but is also important for high-voltage direct current (HVDC) electric power transmission systems, which are currently discussed and planned as part of the "energy transition" in many European countries. In other countries, e.g. China, Brazil and India, huge HVDC traces and grids are already used for the transmission of large energy amounts \cite{hvdc1,hvdc2,hvdc3,hvdc4,hvdc5}.\\
The general approach to measure high voltage is to scale it with a HV divider to a range, where it can be compared to a reference voltage source\footnote{E.g. a Fluke 732A 10~V reference voltage source.}, which is calibrated by a metrology laboratory like the German National Metrology Institute Physikalisch-Technische Bundesanstalt (PTB) with a Josephson voltage standard \cite{paper_josephson}.\\
Precision HV dividers to the ppm\footnote{Parts per million, 1 ppm = $1\cdot 10^{-6}$}-level are commercially available only for voltages up to 1~kV. One key problem for the operation of ppm-precise HV dividers for higher voltages is the lack of traceable calibration methods with the required precision. HV dividers are composed of resistors and therefore generally show a voltage- and time dependent behavior. This is mainly caused by thermal loads and leakage currents with respect to different voltage ranges and powers. Hence, calibration values obtained at low voltages in the order of 1~kV can not be extrapolated for higher voltages without corrections.\\
Up to now the only possibility to calibrate a HV divider to the ppm-level is to transport the unit under test to a metrology center and compare it to a well-known standard HV divider like the MT100 \cite{marx_mt100} of PTB for direct voltages up to 100 kV. The voltage dependency of the MT100 is proven at the nominal voltage of each resistor. But the traceable comparison of the entire divider with a known reference is not possible at high voltages. Therefore, the uncertainty budget of the MT100 has a major contribution caused by the linearity extension leading to an overall expanded uncertainty of $2\cdot10^{-6}$.\\
Recently two new methods for an absolute calibration of HV dividers were reported in \cite{noertershaeuser2018} and \cite{k35_kr}, where uncertainties in the range of $5\cdot10^{-6}$ could be achieved. However, these methods require a complex and partially unique experimental set-up (e.g. an ion beamline with a laser spectroscopy set-up or the 70~m long KATRIN neutrino mass experiment), making these methods very difficult to apply in laboratories with only commercially available equipment.\\
In this paper we present a newly developed method for absolute calibrations of HV divider to the ppm-level by measuring a traceable differential voltage under HV conditions, which can be performed with commercially available devices. The next section gives an overview over the basic set-up of HV dividers and their former calibration techniques. Subsequently, the newly developed calibration method will be explained and first measurement results with achieved relative uncertainties of less than $1\cdot10^{-6}$ will be presented.

\section{High-voltage divider characterization}
\label{section_general_hv}
\begin{figure}[t]
	\centering
	\includegraphics[width=0.16\textwidth]{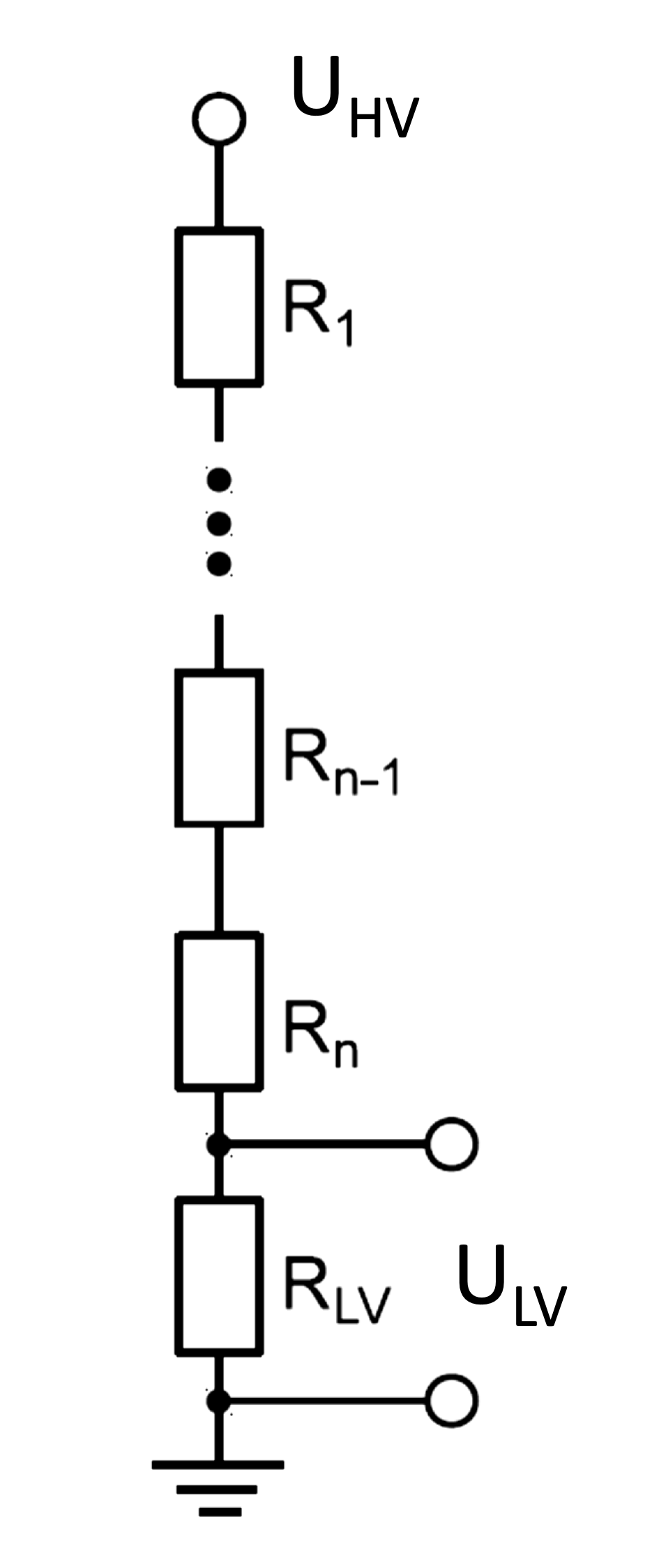}
	\caption{Schematic overview of a simple HV divider. The output voltage $U_{\text{LV}}$ measured over a part $R_{\text{LV}}$ of the resistor chain $R_i$ is proportional to the input voltage $U_{\text{HV}}$. The proportionality factor is called the scale factor $M$.}
	\label{scheme_hv_divider}
\end{figure}

\begin{figure}[t]
	\centering
	\includegraphics[width=0.5\textwidth]{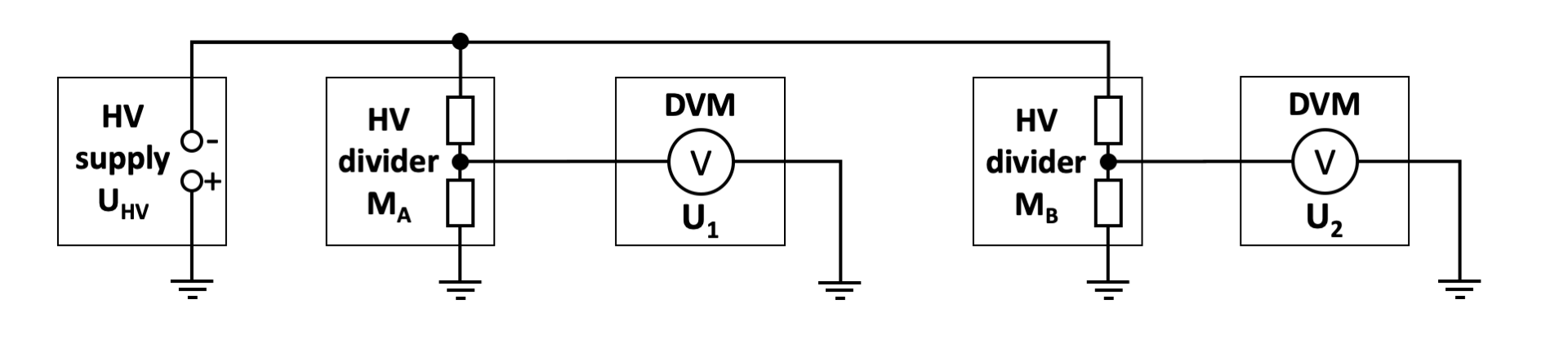}
	\caption{Connection scheme for the calibration of a HV divider with a HV supply ($U_{\text{HV}}$) and a precision DVM to measure the output voltage $U_1$ of the scale factor $M_{\text{A}}$. A reference HV divider with well known scale factor $M_{\text{B}}$ is connected to the same HV source. In combination with a second precision DVM ($U_2$) it is used to determine the input voltage $U_{\text{HV}}$. With commercial equipment this procedure is limited to 1~kV.}
	\label{scheme_lvc_m100}
\end{figure}

\begin{figure}[h]
	\centering
	\includegraphics[width=0.5\textwidth]{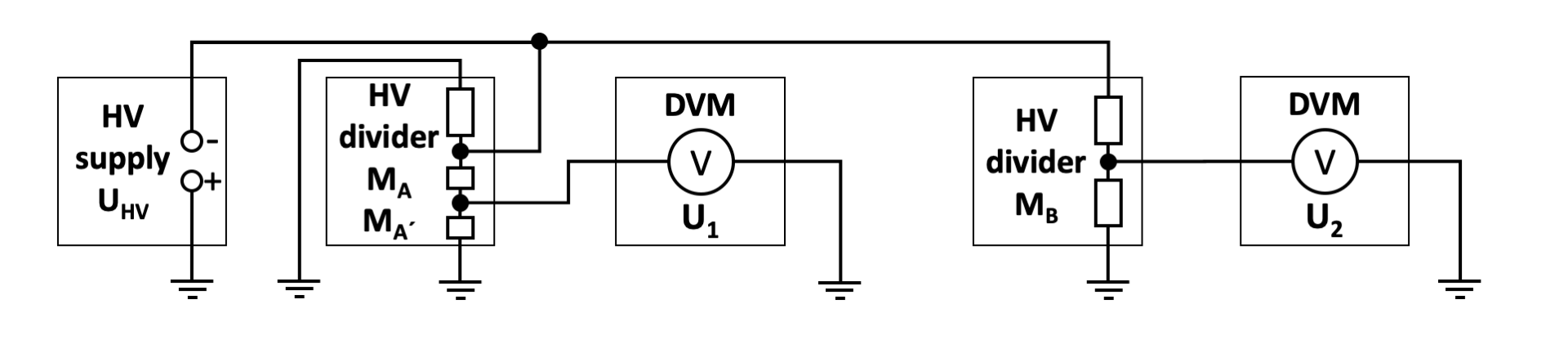}
	\caption{Connection scheme for the calibration of a HV divider with the two scale factors $M_{\text{A}}$ and $M_{\text{A}^{\prime}}>M_{\text{A}}$. Here the voltage created by a HV supply is not connected to the input of the unit under test, but to the scale factor $M_{\text{A}}$ output connection. The scaled voltage $U_1$ is measured with a precision DVM at $M_{\text{A}^{\prime}}$. A reference HV divider with scale factor $M_{\text{B}}$ and a second DVM ($U_2$) are used to determine the input voltage.}
	\label{scheme_lvc_m2000}
\end{figure}

Since high voltages can not be measured directly with ppm-precision, HV dividers are used to scale voltages into the range of typically below 20~V. Here precision digital voltmeters (DVM) are calibrated with 10~V reference sources, which are traceable to a natural standard at metrology institutes.\\
Figure \ref{scheme_hv_divider} shows a schematic overview of a simple HV divider. It consists of a chain of multiple resistors $\sum_{i=1}^n R_{i}$ and a low voltage resistor $R_{\text{LV}}$ connected in series. The output voltage $U_{\text{LV}}$ measured over $R_{\text{LV}}$ is proportional to the input voltage $U_{\text{HV}}$ of the divider. The characteristic observable is the so-called scale factor $M$:
\begin{equation}
	M := \frac{U_{\text{HV}}}{U_{\text{LV}}} = \frac{\sum_{i=1}^n R_{i} + R_{\text{LV}}}{R_{\text{LV}}}
	= \frac{\sum_{i=1}^n R_{i}}{R_{\text{LV}}}+ 1.
	\label{equation_scale_factor}
\end{equation}
Depending on the properties of $R_{\text{LV}}$ compared to the overall resistance, arbitrary and also -- 
if $R_{\text{LV}}$ consists of multiple resistors -- numerous scale factors can be realized.
Following equation (\ref{equation_scale_factor}), $M$ depends on the ratio of $R_{\text{LV}}$ and $\sum_{i=1}^n R_{i}$. If HV is applied to a voltage divider, its individual resistances might change due to dissipated power caused by Joule heating. Since the power of heating $P$ scales quadratically with the current $I$ and linearly with a resistance $R$
\begin{equation}
	P \propto I^2 \cdot R = \frac{U^2}{R},
	\label{equation_power}
\end{equation} 
one can conclude, that the resistances $R_{\text{LV}}$ and $R_{i}$, and thus the scale factor $M$ are voltage dependent:
\begin{equation}
	M = M(U_{\text{HV}}).
\end{equation} 
To mitigate this effect, the total resistance of precision HV dividers is typically in the M$\Omega$-range or higher, limiting the electrical current through the system to less than 1~mA. Furthermore, usually high-quality resistors (e.g. \cite{vishay}) with a low temperature coefficient in a closed stabilized thermal environment are used, resulting in low temperature dependency and long term stability of the scale factor in the (sub)-ppm-range \cite{Thummler:2009rz,Bauer:2013pca,marx_mt100}.\\
In order to calibrate the scale factor $M_{\text{A}}$ of a HV divider, the general procedure is to apply a calibration input voltage $U_{\text{HV}}$ and measure the output voltage $U_1$ with a precision DVM\footnote{In the ideal case the input resistance of a DVM is infinitely high. In reality, the input resistance of the DVM $R_\text{in,DVM}$ (in the 100~G$\Omega$ to 1~T$\Omega$ range for high-end DVM) has to be more than a million times larger than $R_\text{LV}$ to determine the scale factor with ppm-precision. Otherwise the scale factor has to be corrected for $R'_\text{LV} = R_\text{LV} || R_\text{in,DVM}$.}. The input voltage has to be determined with a reference HV divider with well known scale factor $M_{\text{B}}$ and a second precision DVM measuring its output voltage $U_2$:
\begin{equation}
	U_{\text{HV}} = M_{\text{B}} \cdot U_2.
	\label{input_voltage_m100}
\end{equation}
This set-up is shown schematically in figure \ref{scheme_lvc_m100}. 
Following equations (\ref{equation_scale_factor}) and (\ref{input_voltage_m100}), the scale factor of the unit under test can be calculated to be
\begin{equation}
	M_{\text{A}} = \frac{U_2 \cdot M_{\text{B}}}{U_1}.
\end{equation}
Since commercial reference dividers with ppm-precision are only available for voltages up to 1~kV, the calibration with these devices in such a configuration is limited to 1~kV not probing the full range of $M_\text{A}(U_\text{HV})$. Secondly this arangement prefers scale factors of 100:1 or smaller to avoid that the output voltage $U_{\text{1}}$ gets far below the desired 10~V. For example, for a scale factor of 2000:1 the output voltage measured with a DVM would be 0.5~V. Measuring such a small voltage would mean  losing one digit of resolution of the most precise range of the DVM and is therefore not directly traceable to a 10~V reference source used to calibrate the DVM.\\
A standard procedure to avoid this problem is a step-up technique with 1~kV (low voltage) equipment. A prerequisite to apply this method is that the HV divider under test has multiple scale factors, one of them ideally scaling $M_{\text{A}}\approx$ 100:1. In the first step $M_{\text{A}}$ has to be calibrated with the direct method mentioned above with 1~kV. In a second step the higher scale factor $M_{\text{A}^{\prime}}$ is calibrated by applying $U_{\text{HV}}$ not to the regular divider input, but to the $M_{\text{A}}$ output connection. In this arrangement the voltage drop over the low voltage resistors $R_{LV}$ at a calibration voltage $U_{\text{HV}}\leq$~1kV is comparable to the voltage drop over the resistors at an input HV of $U_{\text{HV}}\cdot M_{\text{A}}$. The connection scheme for this calibration method is shown in figure \ref{scheme_lvc_m2000}.
The set-up for the determination of the input voltage is similar to the previous method. For the calculation of $M_{\text{A}^{\prime}}$ one has to multiply the determined input voltage with $M_{\text{A}}$:
\begin{equation}
	M_{\text{A}^{\prime}} =  \frac{U_2 \cdot M_{\text{B}}}{U_1} \cdot M_{\text{A}}.
	\label{equation_for_m_prime_correctly}
\end{equation}
One disadvantage of this method is, that the upper part of the divider with the resistors $R_i$ is not loaded with the correct voltage $M_\text{A} \cdot U_\text{HV}$. This means, that the voltage dependency of the scale factor $M_\text{A}$ is not determined and included in the analysis properly.  For a completely traceable calibration of a HV divider, the voltage dependency of the scale factors has to be taken into account correctly\footnote{As described above, the traceability of the single resistors is possible.}. In order to do so, we developed a novel ppm-precise absolute calibration method for HV dividers, which uses the low voltage equipment described above, elevated on a high-voltage potential.

\section{Novel absolute calibration method} 
\label{section_new_method}
\begin{figure*}[h]
\centering
\includegraphics[width=0.8\textwidth]{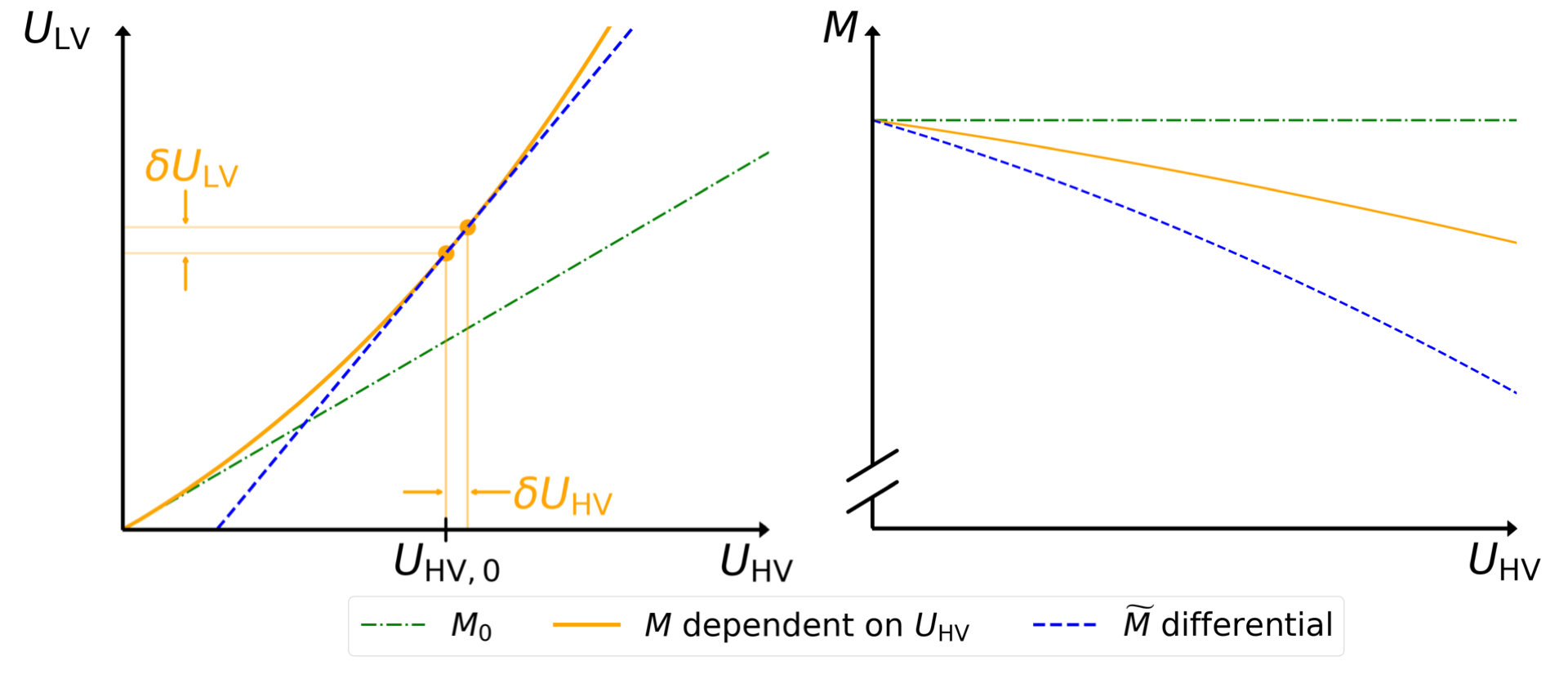}
\caption{Illustration of scale factors as function of the input- and output voltages. Left: Output voltage as function of input voltage. Right: Scale factor as function of input voltage. A constant scale factor appears as a straight line. If the scale factor is dependent on the input voltage (see orange solid line), a deviation from the constant case is observed. For each input voltage $U_\text{HV}$, the differential scale factor is measured as a change of input- and output voltages. This is illustrated at the left at a certain input voltage $U_{\text{HV,0}}$. The differential scale factor $\widetilde M$ appears as slope of the line through the two points $U_{\text{HV,0}}$ and $U_{\text{HV,0}}$+$\delta U_{\text{HV}}$ (blue dashed line). $M_0$ notifies the scale factor derived at $U_\text{HV} \approx 0$ (green dash dotted line).}
\label{python_plots}
\end{figure*}

\begin{figure}[ht]
	\centering
	\includegraphics[width=0.5\textwidth]{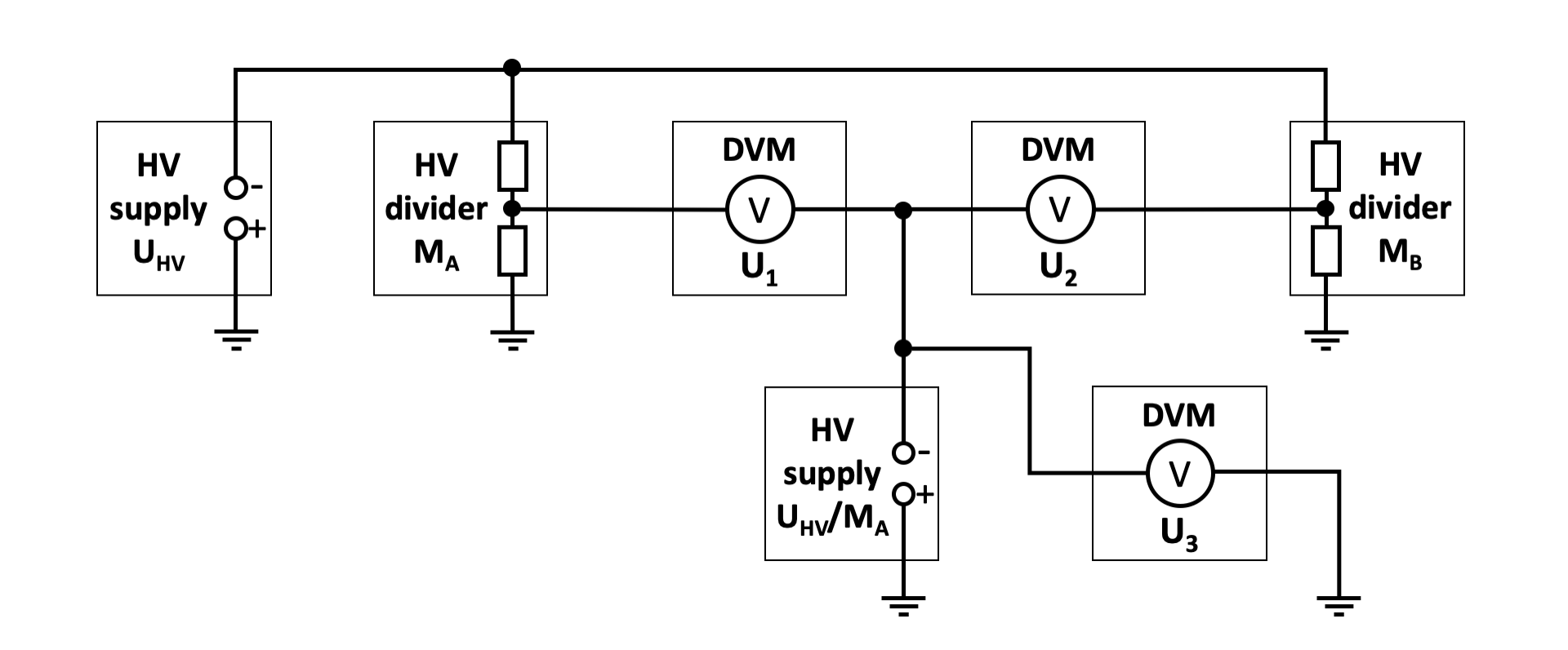}
	\caption{Connection scheme for the measurement of the ratio $\mu$ of the scale factors $M_{\text{A}}$ and $M_{\text{B}}$. A HV $U_\text{HV}$ is connected to both HV dividers and their output voltages are measured with two DVMs versus a counter voltage as a null volt measurement, which is monitored with a third DVM. The counter voltage labeled $U_\text{HV}/M_\text{A}$ is adjusted such that $U_1 \approx 0$.}
	\label{experimental_setup_mu}
\end{figure}

\begin{figure}[ht]
	\centering
	\includegraphics[width=0.5\textwidth]{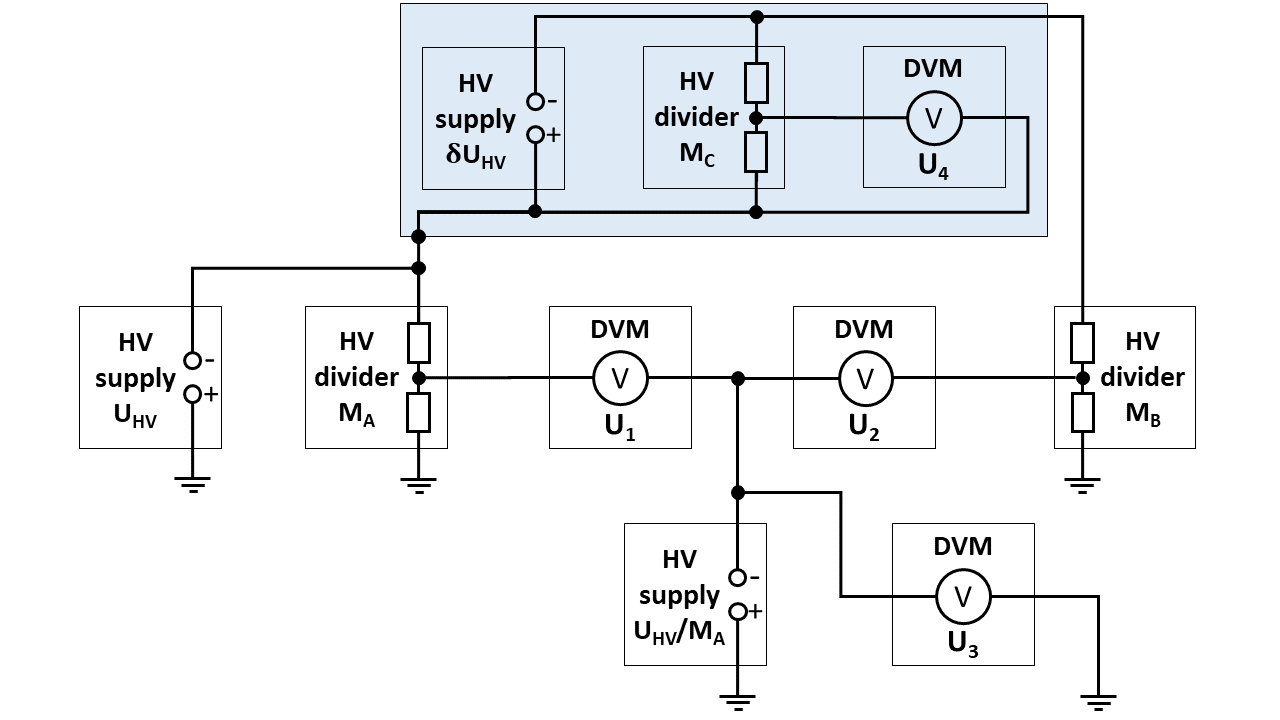}
	\caption{Connection scheme for differential scale factor measurement. On top of a high-voltage potential an additional calibration voltage is created, which is applied to the unit under test. The reference HV divider is unaffected by the calibration voltage. The devices in the blue shaded box are located in a HV cage and read out via an optical link.}
	\label{m_differential_measurement}
\end{figure}

The basic idea of the novel absolute calibration method is to determine the voltage dependency of the scale factors of a HV divider by measuring a differential scale factor directly at high voltages with commercially available equipment. This is especially important for scale factors up to 100:1, since they are used in a step-up technique to calibrate higher scale factors (see section \ref{section_general_hv}).\\
As defined in equation (\ref{equation_scale_factor}) the scale factor is the possibly voltage dependent factor between the input- and output voltage of a HV divider. For a given input voltage the corresponding output voltage can be approximated by a Taylor expansion around $U_\text{HV} = 0$:
\begin{equation}
	U_{\text{LV}} = a \cdot U_{\text{HV}} + b \cdot U_{\text{HV}}^2 + c \cdot U_{\text{HV}}^3 + d \cdot U_{\text{HV}}^4 + ...
	\label{Uout_function_of_Uin}
\end{equation}
with the coefficients $a$, $b$, $c$ and $d$ (neglecting higher orders\footnote{The thermal heat scales with the electric power $P$, which scales with $U^2$ (see equation \ref{equation_power}). The precision resistors of the HV dividers we used have a non-linear, close to quadratic characteristic curve. In addition, we expect a non-linear behavior caused by the thermal control system of our HV dividers and therefore the Taylor expansion may have higher orders, depending on the resistors and the temperature control system. In our measurements the Taylor approximation of second (fourth) order was sufficient for the K65 (G35) HV divider (see section \ref{calibration_results}).}). For the voltage independent case the parameters $b$, $c$ and $d$ are zero and $a$ is the inverse of the constant part of the scale factor $M_0$:
\begin{equation}
	a = \frac{1}{M_{0}}.
	\label{parameter_a}
\end{equation}
For the realistic case of a voltage dependent scale factor we can derive from equation (\ref{equation_scale_factor}) and (\ref{Uout_function_of_Uin}):
\begin{equation}
	M = \frac{1}{a + b \cdot U_{\text{HV}} + c \cdot U_{\text{HV}}^2 + d \cdot U_{\text{HV}}^3}.
	\label{equation_theory_real_scale_factor}
\end{equation}
We define a differential scale factor $\widetilde{M}$ as the derivative of $U_{\text{HV}}$ with respect to $U_{\text{LV}}$ at $U_{\text{HV}}$:
\begin{align}
    	\widetilde{M} = \frac{\delta U_{\text{HV}}}{\delta U_{\text{LV}}}\bigg|_{U_{\text{HV}}} = \frac{1}{\frac{\partial U_{\text{LV}}}{\partial U_{\text{HV}}}\bigg|_{U_{\text{HV}}}} = \\
	\frac{1}{a + 2 \cdot b \cdot U_{\text{HV}} + 3 \cdot c \cdot U_{\text{HV}}^2 + 4 \cdot d \cdot U_{\text{HV}}^3}.
	\label{equation_theory_differential_scale_factor}
\end{align}
The measurement of $\widetilde{M}$ at $U_{\text{HV}}$ is done with the following procedure: at certain input voltages we increase $U_{\text{HV}}$ by a small amount of $\delta U_\text{HV}$ and measure the change of the output voltage $\delta U_\text{LV}$. In the ideal case the voltage increase $\delta U_{\text{HV}}$ is infinitesimal small in order to determine the slope of the scale factor curve at $U_{\text{HV}}$. However, due to technical limitations and because of the ambition to trace the voltage measurement back to a 10~V reference, this is not possible. Hence, we increase the voltage by $\delta U_\text{HV} = 1$~kV, which can be measured with traceable equipment with ppm-precison. Therefore we assume, that the determined scale factor is valid for the input voltage $U_{\text{HV}}~+~\delta U_{\text{HV}}/2$. The two cases of the constant and voltage dependent scale factor are sketched in figure \ref{python_plots}. Additionally $\widetilde{M}$ is illustrated for an exemplary input voltage $U_\text{HV,0}$.
By measuring the differential scale factor for different input voltages the coefficients $a$, $b$, $c$ and $d$ can be determined and used to calculate the scale factor $M$ for any given input voltage. \\
The measurement of $\widetilde{M}$ is split into two steps: figure \ref{experimental_setup_mu} shows the experimental set-up for the first step.
A high voltage $U_{\text{HV}}$ is connected to the HV divider whose scale factor $M_{\text{B}}$ is to be calibrated. Its output voltage $U_2$ is measured with a precision DVM versus a very stable counter voltage\footnote{The ppm-stable counter voltage is provided by a Fluke Calibrator 5720A.} $U_{\text{HV}}/M_{\text{A}}$ as a null volt measurement. By using a counter voltage instead of a measurement versus ground potential it is ensured, that the measured voltage is below 20~V, which can be traced back to a 10~V reference source. The counter voltage is either directly monitored with a third DVM\footnote{Since $U_3$ has to be very stable but does not need to be known such precisely we monitored this voltage with a 6.5 digit DVM of type Fluke 8846A.} ($U_3$) or converted via a reference divider\footnote{For this purpose we used a Fluke reference divider of type 752A.} into the 0 to 20~V range. Additionally a second HV divider ($M_{\text{A}}$) is needed as reference for the unit under test, which is connected to the same HV source. The output voltage of the reference HV divider is also measured with a DVM ($U_1$) versus the counter voltage. In this measurement the ratio of the scale factors $\mu$
\begin{equation}
	\mu := \frac{M_{\text{A}}}{M_{\text{B}}} = \frac{U_2 + U_3}{U_1 + U3} \approx 1 + \frac{U_2}{U_3}
	\label{equation:mu_ratio}
\end{equation}
can be determined applying Kirchhoff$^{\prime}$s circuit laws. The approximation on the right of equation (\ref{equation:mu_ratio}) is only valid for $U_1 \approx 0$ and should only illustrate that $\mu$ does not require a precise determination of $U_3$. This counter voltage is a key to achieve the ppm-precision for the novel absolute calibration method.
The ratio $\mu$ can be measured with a short-term precision of the order of below $10^{-7}$ without knowing the single scale factors $M_{\text{A}}$ and $M_{\text{B}}$, since it only depends on the measured voltages $U_{1,2,3}$, which are determined with precision DVMs. Since both null volt measurements $U_{1}$ and $U_{2}$ are measured with the same counter voltage, both scale factors have to be of similar magnitude in order to not exceed the 20~V range of the DVM.\\
In the second step the input voltage of the HV divider under test is increased by $\delta U_{\text{HV}}$, which is generated and measured on top of the HV potential $U_{\text{HV}}$ (see figure \ref{m_differential_measurement}). The input voltage of the reference HV divider stays constant as well as the counter voltage, any potential change would be detected by continuously measuring $U_1$ and $U_3$. The DVM, which is used to measure the output voltage of the divider under test, will measure a voltage increase of $\delta U_{\text{HV}}/\widetilde M_{\text{B}}$. According to Kirchhoff$^{\prime}$s circuit- and Ohm$^{\prime}$s laws the differential scale factor is given by
\begin{equation}
	\widetilde{M}_{\text{B}} = \frac{U_1 \cdot M_{\text{A}} + U_4 \cdot M_{\text{C}}}{U_2 + (1 - \mu) \cdot U_3}.
	\label{equation_differential_scale_factor}
\end{equation}
As denoted in equation (\ref{equation_differential_scale_factor}) the scale factor of the reference HV divider $M_\text{A}$ is needed to calculate $\widetilde{M}_{\text{B}}$. However, the term $U_1 \cdot M_{\text{A}}$ is close to zero since $U_1$ is a null volt measurement against the stable counter voltage adjusted to $U_1 \approx 0$. Hence, the dominant factor of the numerator is $U_4 \cdot M_{\text{C}}$, which means, that the absolute value of $M_{\text{A}}$ needs to be stable but does not have to be known precisely in order to calibrate the unit under test to the ppm-level. The measurements, which are presented in the next section, showed, that an uncertainty of up to $1\cdot10^{-4}$ can be allowed for $M_{\text{A}}$, without changing the calibration result for $M_{\text{B}}$ on the $1\cdot10^{-7}$ level.
Secondly, the uncertainty of $U_3$ is not important since the ratio of the scale factors $\mu$ is close to 1. Therefore $U_2$ and its uncertainty are dominating the denominator for the determination of $\widetilde{M}_{\text{B}}$.

\section{Calibration results}
\label{calibration_results}
\begin{figure}[t]
	\centering
	\includegraphics[width=0.15\textwidth]{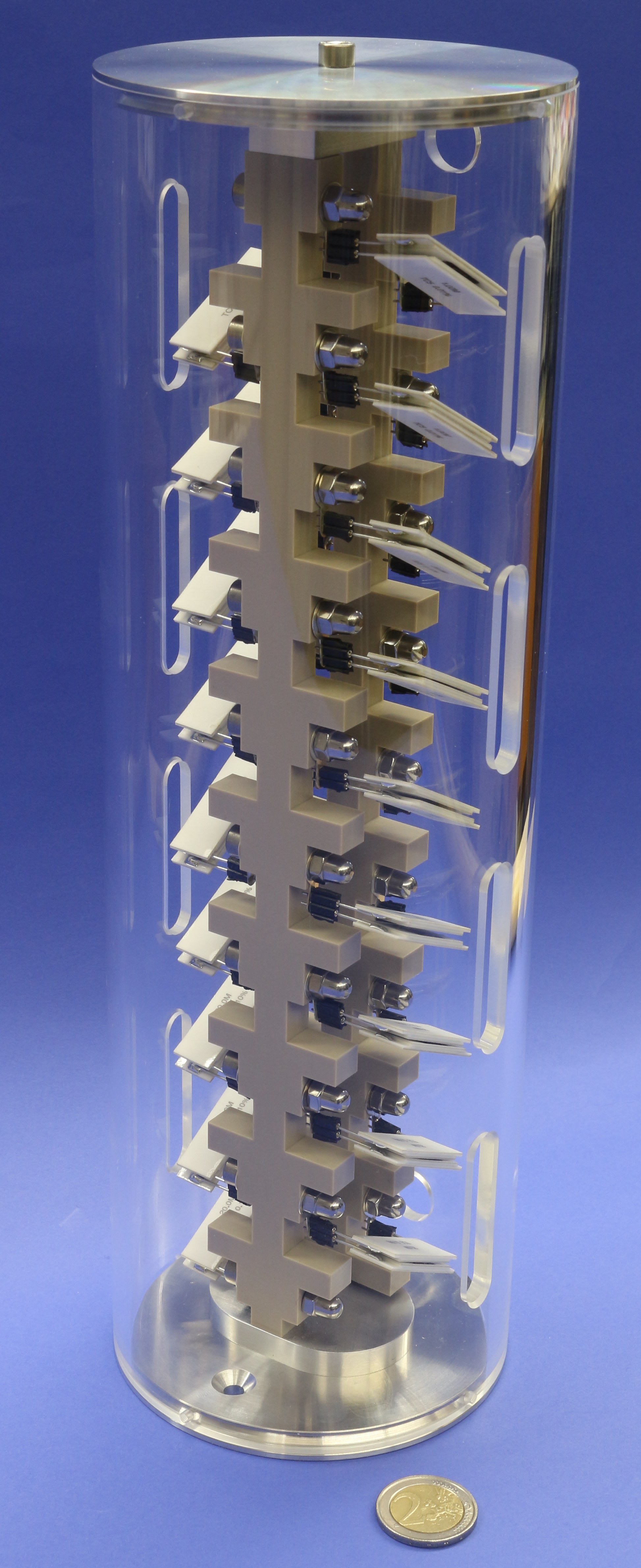}
	\caption{Picture of custom made HV divider consisting of two times 18 precision resistors (30 x 20 M$\Omega$ Caddock type USF 371 and 6 x 1 M$\Omega$ Caddock type USF 370) connected in series. The scale factor $M_\text{A}\approx 100$ has a relative uncertainty of about $1\cdot10^{-5}$.}
	\label{caddock_divider}
\end{figure}

\begin{figure}[b]
	\centering
	\includegraphics[width=0.5\textwidth]{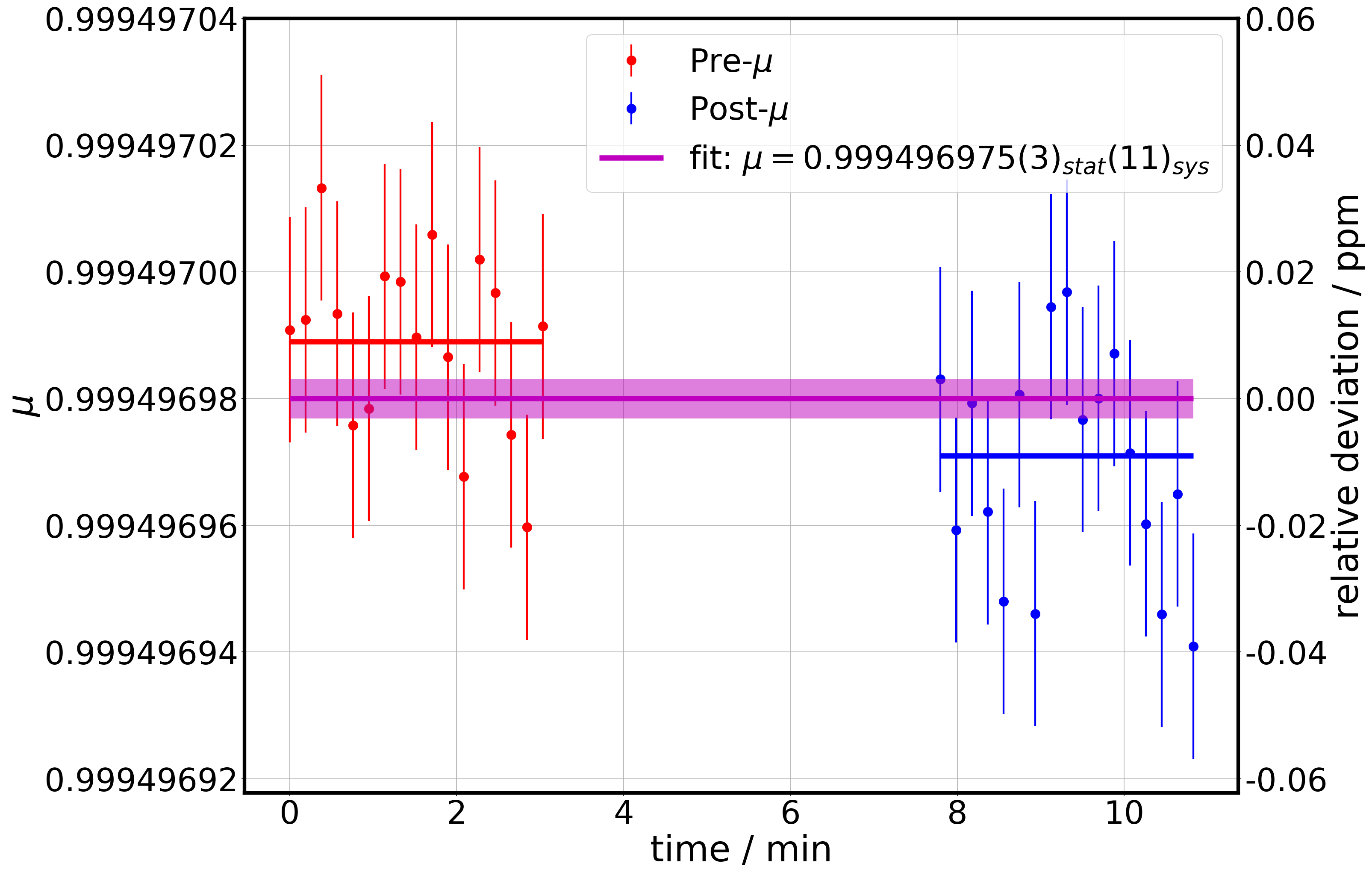}
	\caption{Exemplary measurement of the scale factor ratio $\mu$ of the unit under test and a reference HV divider measured at $U_{\text{HV}}$~=~-18.6~kV. As both scale factors are about 100:1, the ratio is close to one. The data has been fitted with a constant in order to determine the mean value. We did not use a polynomial of first order because of the smallness of the effect ($1\cdot 10^{-8}$ level). Since for short time intervals only transfer uncertainties are known, which are valid for 20 min, we use the measured fluctuations in order to determine the statistical uncertainties for longer periods. Therefore the error bars are scaled such, that the quadratic deviation per number of degrees of freedom is equal to one ($\chi^2_r = 1$). For the systematic uncertainties we determined the 24~h uncertainties of each DVM with a reference voltage source.}
	\label{single_plot_mu}
\end{figure}

\begin{figure}[t]
	\centering
	\includegraphics[width=0.5\textwidth]{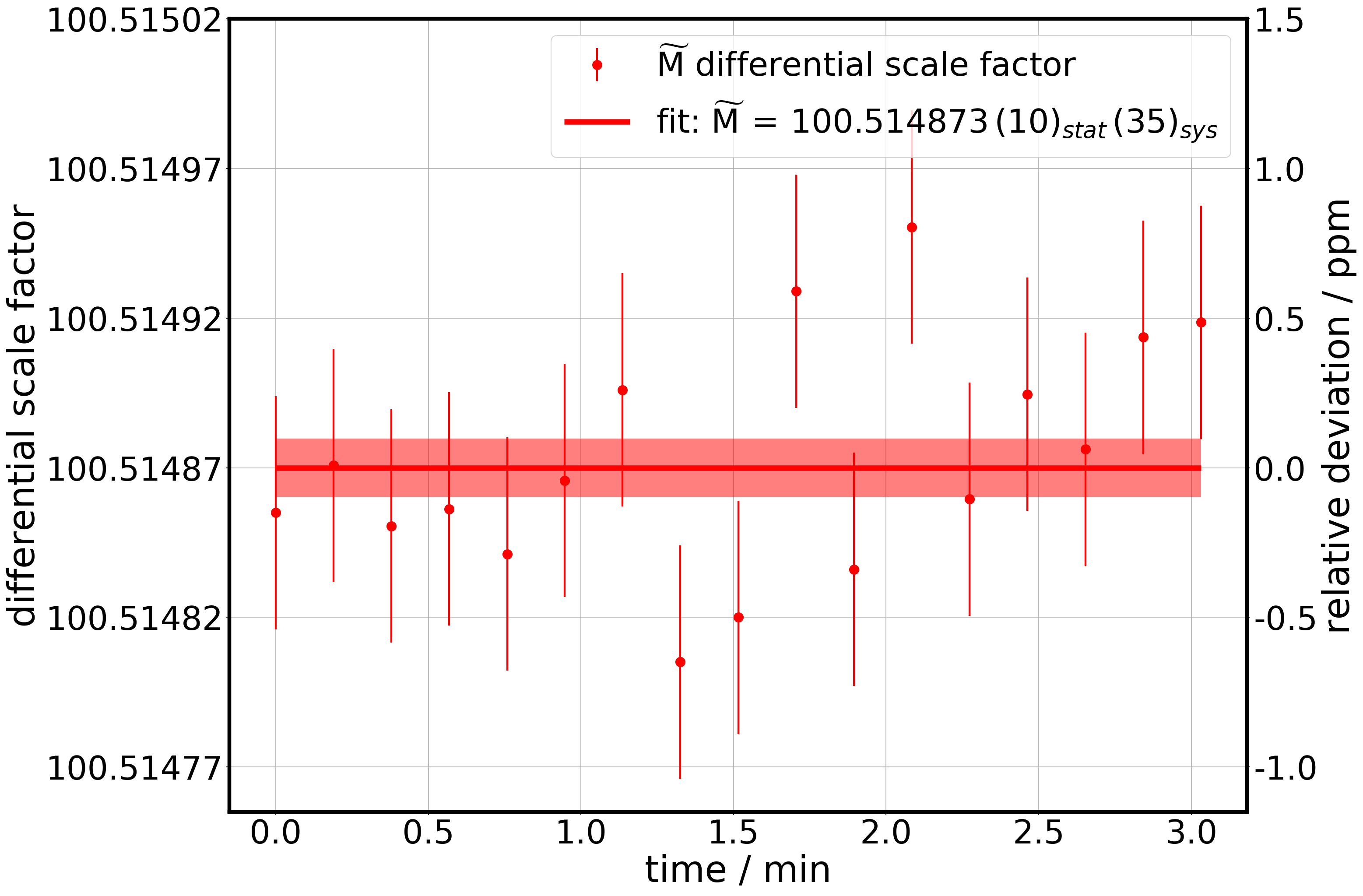}
	\caption{Exemplary measurement of differential scale factor determined with the newly developed absolute calibration method measured at $U_{\text{HV}}$~=~-18.6~kV. The data has been fit with a constant in order to determine the mean value. Since for short time intervals only transfer uncertainties are known, which are valid for 20 min, we use the measured fluctuations in order to determine the statistical uncertainties for longer periods. Therefore the error bars are scaled such, that the quadratic deviation per number of degrees of freedom is equal to one ($\chi^2_r = 1$). For the systematic uncertainties we determined the 24~h uncertainties of each DVM with a reference voltage source.
	}
	\label{single_plot_m_differential}
\end{figure}

During a measurement campaign in early 2018 numerous calibrations of different HV dividers have been performed. The main goal was to check the reproducibility and long-term stability of the newly developed absolute calibration method as well as its capability to measure the voltage dependency of scale factors. 
The measurements were performed with two ppm-precise HV dividers K65 \cite{Bauer:2013pca} and G35 \cite{g35_paper}, which were also used as reference mutually to crosscheck the results. In addition we built a HV divider with precision resistors \cite{caddock} with a scale factor $M_\text{A}\approx 100:1$ and a relative uncertainty of in the order of $1\cdot10^{-5}$, which was used as reference unit (see figure \ref{caddock_divider}). 
Commercial HV dividers\footnote{We used Fluke 752A reference dividers, which were calibrated on each measurement day before the novel calibration procedure.} were used to measure the calibration voltage $\delta U_{\text{HV}}$ up to 1~kV.
The voltage measurements were performed with 8.5 digit precision DVM\footnote{For measuring $U_1$, $U_2$ and $U_4$ we used the devices Fluke 8508A, Agilent 3458A and Keysight 3458A. The less critical voltage $U_3$ was monitored with a 6.5 digit DVM of type Fluke 8846A.}. Our HV source $U_{\text{HV}}$ and HV divider G35 were limited to 35~kV.\\
As described in section \ref{section_new_method} the stability of the ratio-measurement of the scale factors has been investigated. Figure \ref{single_plot_mu} shows a single $\mu$ determination run consisting of 17 measurements before and 17 measurements after the determination of the differential scale factor $\widetilde M_\text{B}$. In order to determine its mean value, which according to equation (\ref{equation_differential_scale_factor}) is needed to calculate the differential scale factor, the data has been fitted with a constant. As described in the previous section the ratio can be determined without knowing the individual scale factors of both dividers with relative uncertainties smaller than $1\cdot10^{-7}$.
Subsequently $\widetilde{M}_\text{B}$ has been measured according to figure \ref{m_differential_measurement}. The differential scale factor was derived with equation (\ref{equation_differential_scale_factor}), including the calculated mean $\mu$-value determined directly before and after the calibration measurement. Figure \ref{single_plot_m_differential} shows a single measurement of the differential scale factor. The standard deviation is below $5\cdot10^{-7}$. The differential scale factor, always together with the ratio $\mu$, has been measured multiple times each day during the calibration campaign at different voltages. They agreed very well within uncertainties.

\begin{figure}[t]
	\centering 
	\includegraphics[width=0.5\textwidth]{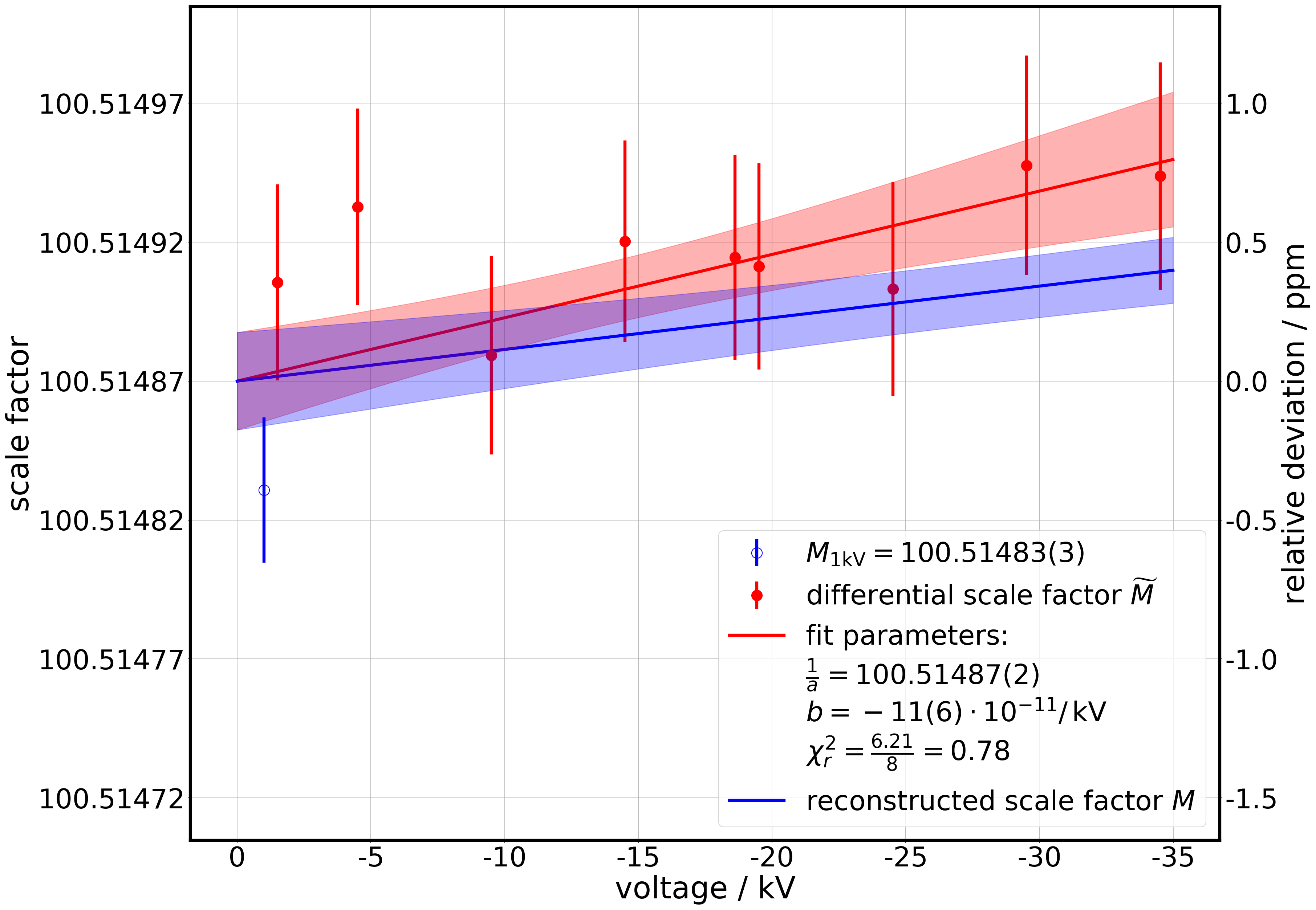}
	\caption{Voltage dependency of the K65 100:1 scale factor determined with the newly developed absolute calibration method. The differential scale factors $\widetilde M$ measured at different voltages (red points) and the low voltage scale factor $M_{\text{1kV}}$ (blue point) are fitted with a polynomial of first order (red line). The error-bars include the statistical and systematic uncertainties. The obtained coefficients are used to calculate the real scale factor $M$ for a voltage range from 0 to 35~kV (blue line).}
	\label{calibration_result_k65}
\end{figure}

\begin{figure}[t]
	\centering 
	\includegraphics[width=0.5\textwidth]{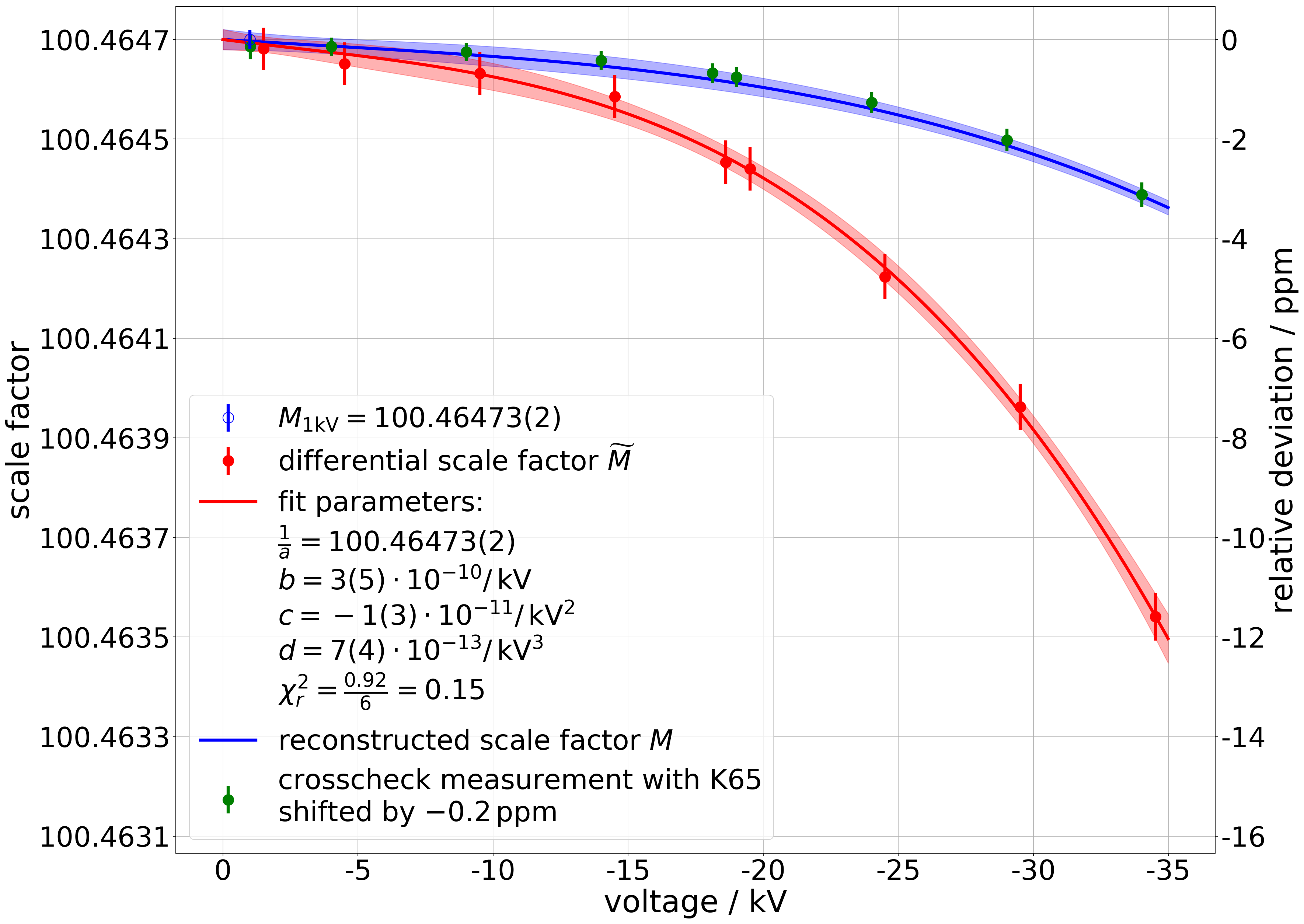}
	\caption{Voltage dependency of the G35 100:1 scale factor determined with the newly developed absolute calibration method. The differential scale factors $\widetilde M$ measured at different voltages (red points) and the low voltage scale factor $M_{\text{1kV}}$ (blue point) are fitted with a polynomial of third order (red line). The error-bars include the statistical and systematic uncertainties. The obtained coefficients are used to calculate the real scale factor $M$ for a voltage range from 0 to 35~kV (blue line). In order to verify the result for the G35, the K65 was used to crosscheck the voltage dependency more than two months later (green points). Note that all green datapoints are shifted by $-2\cdot10^{-7}$ in y direction (see text).}
	\label{calibration_result_g35}
\end{figure}

In order to derive the real scale factor $M_{\text{B}}$ from $\widetilde{M}_{\text{B}}$ we measured the differential scale factor for different voltages up to 35~kV (see figure \ref{calibration_result_k65} and \ref{calibration_result_g35}) and fitted the data\footnote{The data was fitted with MINUIT \cite{refMINUIT}} according to equation (\ref{equation_theory_differential_scale_factor}) to obtain the coefficients  $a, \ b \ c \text{ and } d$. We also included the low voltage calibration values measured as described in section \ref{section_general_hv} (see set-up in figure \ref{scheme_lvc_m100}) into the analysis. Since in that measurements the real scale factor is determined, we used a combined fit to describe all data points\footnote{The fit function is a sum of equations (\ref{equation_theory_real_scale_factor}) for the data point obtained with the low voltage calibration measurement and (\ref{equation_theory_differential_scale_factor}) for the data points of the differential scale factor determination.}. Subsequently $M_{\text{B}}$ is calculated using equation (\ref{equation_theory_real_scale_factor}). For the K65 HV divider a negligible linearity below  $1\cdot10^{-6}$ over the whole input range was observed, which is within the uncertainties in agreement with former calibration measurements at PTB \cite{Bauer:2013pca}. Here a linear voltage dependency ($c=0=d$) was assumed for the fit, as indicated by $\chi^2$-studies of higher orders.\\
The scale factor $M_{\text{B}}$ derived this way for the G35 HV divider showed deviations of up to $3.3\cdot10^{-6}$ at -35~kV compared to the low voltage scale factor $M_{\text{1kV}}$. We crosschecked this by comparing the scale factor $M_{\text{B}}$ of G35 with the one measured directly with the help of K65 using a set-up as shown in figure \ref{scheme_lvc_m100} two months later. Thus, we could confirm the result obtained for the linearity measurement with the novel absolute calibration method. To get an excellent agreement the absolute value of the scale factor required a constant offset of $-2\cdot10^{-7}$ over the full range of -35~kV. This shift exceeds the combined short-term uncertainties (voltage dependent, average about $1\cdot10^{-7}$) for the real scale factor. However, we consider an additional relative uncertainty of $\pm$ $5\cdot10^{-7}$ for the absolute value of the scale factor to be realistic, since all previous low voltage- and high-voltage measurements showed this level of uncertainty, when repeated later on a time scale of weeks or months. Therefore it is reasonable to shift data points of measurements with a significant time difference (here more than 2 months for the comparison shown in figure \ref{calibration_result_g35}) with a constant offset, in order to check the voltage dependency. For future HV measurements with the G35 the ppm-precise voltage-dependent scale factor obtained with the presented work in this article can be used considering the corresponding uncertainties.

\begin{figure}[t]
	\centering 
	\includegraphics[width=0.5\textwidth]{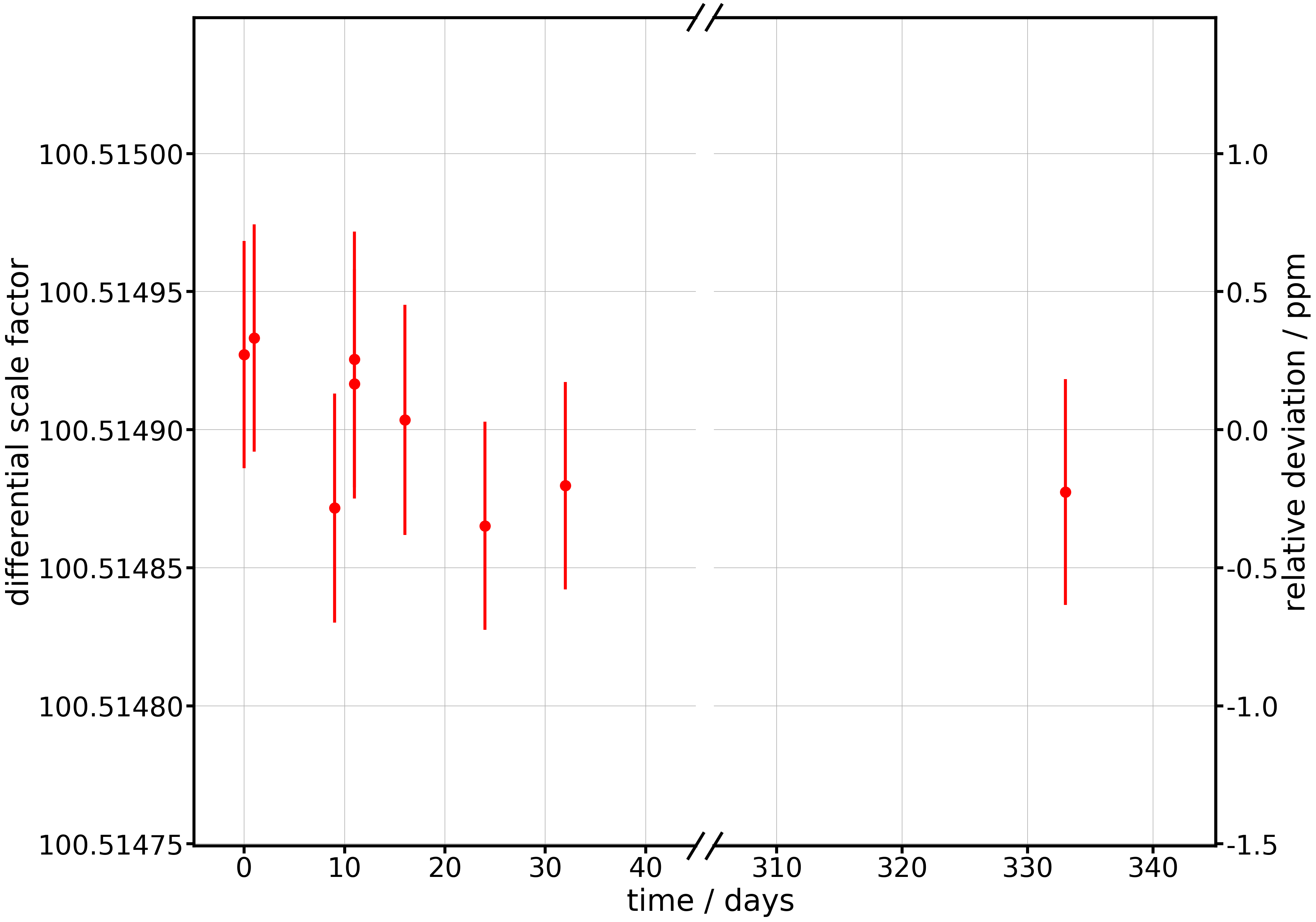}
	\caption{Differential scale factor $\widetilde M$ of the K65 measured at a voltage of $U_{\text{HV}} = -18.6$ kV. The  error-bars  include  the  statistical  and  systematic uncertainties. Over a time period 333 days all measurements of the differential scale factor show a scattering  below $5\cdot10^{-7}$.}
	\label{diff_scale_time}
\end{figure}

We investigated also the long term stability of $\widetilde M$. Figure \ref{diff_scale_time} shows the differential scale factor of the K65 HV divider measured over a time period of about 330 days. The scattering of the determined values of $\widetilde M$ is below $\pm$~$5\cdot10^{-7}$. Compared to the stability of the K65 of $2\cdot10^{-8}$ per month determined at PTB in 2011 (for the 100:1 scale factor), the results obtained with the newly developed absolute calibration technique are in good agreement, confirming the general principle and functionality of this method.

\begin{figure}[h]
	\centering 
	\includegraphics[width=0.5\textwidth]{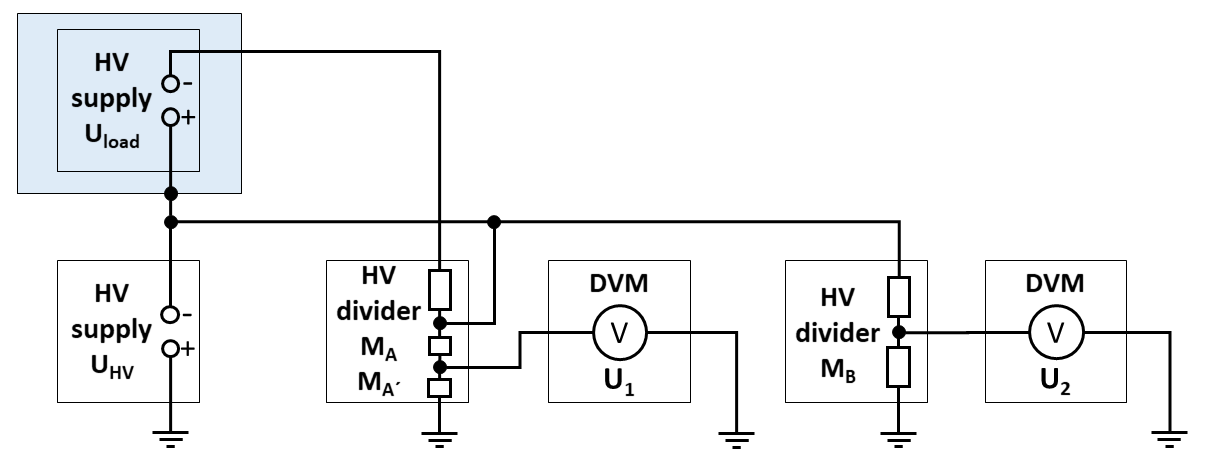}
	\caption{Connection scheme for the corrected determination of $M_{\text{A}^{\prime}}$. The input voltage $U_{\text{HV}}$ is connected to the scale factor output $M_{\text{A}}$ of the unit under test. The upper part of the HV divider with the resistors $R_{\text{i}}$ is loaded with the voltage $U_{\text{load}}=U_{\text{HV}}\cdot M_{\text{A}}$ created by an additional HV supply, which is operated on the potential of $U_{\text{HV}}$ in a HV cage. A second HV divider with the well known scale factor $M_{\text{B}}$ is used to determine $U_{\text{HV}}$.}
	\label{m2000_loaded}
\end{figure}

For the calibration of scale factors $M_{\text{A}^{\prime}}$ $>$~100:1 the procedure similar to the one described in figure \ref{scheme_lvc_m2000} can be used, but to load the resistors $R_i$ correctly, the corresponding HV is additionally given to the input of the HV divider under calibration using a HV cage (see figure \ref{m2000_loaded}).
The wanted scale factor $M_{\text{A}^{\prime}}$ can be calculated
according to equation (\ref{equation_for_m_prime_correctly}). The critical scale factor  $M_\text{A}~\leq~100$ is determined with the novel absolute calibration method. Thus, the issues regarding traceability and the previously neglected voltage dependencies of $M_\text{A}$ and $M^\prime_\text{A}$ vanish. \\
Our estimated uncertainty budget for the differential scale factor is shown in table \ref{table_uncertainty_budget}.
\begin{table*}[]
\caption{Estimated uncertainty budget for the systematic uncertainty of the differential scale factor with most important contributions (shown for an exemplary measurement). For all parameter values $p$ we considered a Gaussian distribution (1 $\sigma$) for the uncertainty $\Delta p$ (see section \ref{calibration_results} for details about the used devices and their uncertainties). The contribution of each parameter is the product of the sensitivity coefficient ($\frac{\partial{\widetilde{M}}}{\partial{p}}$) and $\Delta p$. The relative importance of each contribution is calculated by $\frac{(\frac{\partial{\widetilde{M}}}{\partial{p}} \cdot \Delta p)^2}{(\Delta\widetilde{M}_{tot})^2 }$.}
\resizebox{\textwidth}{!}{%
\begin{tabular}{@{}lllllll@{}}
\hline
\toprule
Parameter                                                              & value p     & abs. uncertainty & unit & sensitivity coeff. & contribution & rel. importance (\%) \\ \midrule
\hline
$M_\mathrm{C}$ HV divider (see fig. \ref{m_differential_measurement})                                             & 100.000000  & 0.000017         &     & 1.01               & 0.000017     & 22.59                \\
$U_4$ DVM (cal. with 10 V ref.)                                        & -10.0000928 & 0.0000012        & V    & 10.05              & 0.000012     & 11.32                \\
$U_4$ DVM ($\widetilde{M}$, see fig. \ref{m_differential_measurement}) & -10.0027489 & 0.0000012        & V    & -10.05             & -0.000012    & 11.32                \\
$U_2$ DVM ($\mu$, see fig. \ref{experimental_setup_mu})                & 0.0935306   & 0.0000011        & V    & -10.10             & -0.000012    & 10.67                \\
$U_2$ DVM ($\widetilde{M}$, see fig. \ref{m_differential_measurement}) & -9.8580844  & 0.0000011        & V    & 10.10              & 0.000012     & 10.67                \\
$U_2$ DVM (cal. offset)                                                & -0.0000067  & 0.0000011        & V    & 10.05              & 0.000011     & 10.57                \\
$U_2$ DVM (cal. with 10 V ref.)                                        & -10.0000948 & 0.0000011        & V    & -10.05             & -0.000011    & 10.57                \\
\midrule
\hline
other uncertainties                                                    &             &                  &      &                    & 0.000012     & 12.28                \\
\midrule
\hline
total uncertainty                                                   & 100.514876  & 0.000035         &     &                    &              & 100                  \\ \bottomrule
\hline
\end{tabular}%
}
\label{table_uncertainty_budget}
\end{table*}
The overall relative uncertainties of about $4\cdot10^{-7}$ are mainly dominated by the two devices, which are operated on the HV potential (about 50~\%): the 1~kV reference divider and the corresponding DVM. Accordingly their calibration before the measurement is of crucial importance. Furthermore at this level of precision also the resistances of the cabling becomes relevant. Especially on the HV side of the set-up cable resistances, which can be in the order of 1~$\Omega$, can influence the calibration result when they are not included in the analysis\footnote{The used reference HV divider Fluke 752A has an input resistance of 2~M$\Omega$. This means, that a cable resistance in the order of 1~$\Omega$ can influence the calibration result on the ppm-level. However, this is more important for the low voltage calibration described in section \ref{section_general_hv}, since the effect nearly cancels out in the  two steps of the differential scale factor measurement of the novel absolute calibration method.}. Finally, as described above, an additional uncertainty of about $5\cdot10^{-7}$ for the absolute value of the scale factor has to be assumed.

\section{Conclusion}
Precision measurements of DC high voltages are important for different applications in fundamental research and applied sciences. In order to measure HV to the ppm-level precision HV dividers are used to scale the voltage into ranges below 20~V, where they  can be compared to voltage references traceable to natural standards at metrology institutes. The scale factors of HV dividers usually are voltage- and time dependent and have to be calibrated regularly. Former calibration methods could only consider this by extrapolating the voltage dependency of individual resistors. In this work we presented a newly developed absolute calibration method for HV dividers, which overcomes this issues and allows a traceable calibration by determining a differential scale factor measured directly at high voltages. We have shown that the systematic uncertainty is in the order of less than $1\cdot10^{-6}$. This method can be performed with commercially available equipment and therefore is not restricted to metrology institutes, but offers measurements of linearities of HV dividers with ppm-precision for a wide range of applications. A comparison of this work and other, recently developed calibration techniques is given in \cite{abs_cal_compare}.\\
There are also investigations to apply this method in order to measure the linearity behaviour of precision compressed gas HV capacitors.

\section*{Acknowledgements}
This work was supported by Ministry for Education and Research BMBF (05A14PMA and 05A17PM3). We would like to express our special thanks to R. Zirpel and C. Rohrig of PTB for the calibration of our 10~V reference sources. We gratefully acknowledge the support from C. Huhmann and H.-W. Ortjohann of the Institut f{\"u}r Kernphysik at M{\"u}nster University during the set-up and commissioning of the HV cage and G35 HV divider.

\section*{References}
\bibliographystyle{spphys}

\bibliography{biblio}

\end{document}